\documentclass[aps,prl,reprint,twocolumn,superscriptaddress,floatfix,nofootinbib,longbibliography]{revtex4-2}
\usepackage{graphicx,amsmath,amsfonts,amssymb,amsthm,xr}
\usepackage{epsfig,amsmath,amssymb,color,dsfont,upgreek,physics}
\usepackage{mathrsfs}
\usepackage{mathtools}
\usepackage{bbold}
\usepackage{float}
\usepackage[caption=false]{subfig}

\usepackage[bookmarks=true,colorlinks,linkcolor=OrangeRed,urlcolor=NavyBlue,citecolor=RoyalBlue]{hyperref}
\usepackage[dvipsnames]{xcolor}
\usepackage{orcidlink}

\usepackage{graphicx}
\usepackage{dcolumn}
\usepackage{bm}
\usepackage{color}

\newcommand{\Zt}{\mathbb{Z}_2}
\newcommand{\rr}{\mathbf{r}}

\begin{document}

\preprint{}

\title{\textbf{String Breaking in a $2+1$D $\mathbb{Z}_2$ Lattice Gauge Theory} }

\author{Umberto Borla${}^{\orcidlink{0000-0002-4224-5335}}$}
\affiliation{Racah Institute of Physics, The Hebrew University of Jerusalem, Givat Ram, Jerusalem 91904, Israel}

\author{Jesse J.~Osborne${}^{\orcidlink{0000-0003-0415-0690}}$}
\affiliation{School of Mathematics and Physics, The University of Queensland, St.~Lucia, QLD 4072, Australia}

\author{Sergej Moroz${}^{\orcidlink{0000-0002-4615-2507}}$}
\affiliation{Department of Engineering and Physics, Karlstad University, Karlstad, Sweden}
\affiliation{Nordita, Stockholm University and KTH Royal Institute of Technology, 10691 Stockholm, Sweden}

\author{Jad C.~Halimeh${}^{\orcidlink{0000-0002-0659-7990}}$}
\email{jad.halimeh@physik.lmu.de}
\affiliation{Max Planck Institute of Quantum Optics, 85748 Garching, Germany}
\affiliation{Department of Physics and Arnold Sommerfeld Center for Theoretical Physics (ASC), Ludwig Maximilian University of Munich, 80333 Munich, Germany}
\affiliation{Munich Center for Quantum Science and Technology (MCQST), 80799 Munich, Germany}

\date{\today}

\begin{abstract}
String breaking is an intriguing phenomenon crucial to the understanding of lattice gauge theories (LGTs), with strong relevance to both condensed matter and high-energy physics (HEP). Recent experiments investigating string breaking in $2+1$D (two spatial and one temporal dimensions) LGTs motivate a thorough analysis of its underlying mechanisms. Here, we perform matrix product state (MPS) simulations of string breaking in an experimentally relevant $2+1$D $\mathbb{Z}_2$ LGT in the presence of two external charges. We provide a detailed	 description of the system in the confined phase, highlight a number of mechanisms which are responsible for string breaking, and argue that magnetic fluctuations have a stabilizing effect on the strings. Moreover, we show that deep in the confined regime the problem is dual to one-dimensional free fermions hopping on an open chain. Our work elucidates the microscopic processes of string breaking in $2+1$D LGTs, and our findings can be probed on current superconducting-qubit quantum computers.
\end{abstract}

\maketitle

\textbf{\emph{Introduction.---}}Gauge theories lie at the heart of the Standard Model of particle physics \cite{Weinberg_book}. They describe the interactions between elementary particles as mediated by gauge bosons \cite{Gattringer_book}, and are used to analyze measurements at dedicated particle colliders \cite{Ellis_book}. Their lattice manifestation, LGTs, have enabled non-perturbative calculations of quantum chromodynamics (QCD), giving insights into the nature of quark confinement \cite{Wilson1974}. Although initially formulated for high-energy physics (HEP), their usefulness has extended to condensed matter \cite{Wegner1971,Kogut_review,wen2004quantum} as well as quantum many-body physics \cite{Bernien2017,Surace2020,Su2022,Desaules2022weak,Desaules2022prominent}. Recently, there has also been a concerted effort to realize LGTs on quantum hardware \cite{Martinez2016,Klco2018,Goerg2019,Schweizer2019,Mil2020,Yang2020,Wang2021,Zhou2022,Wang2023,Zhang2023,zhu2024probingfalsevacuumdecay}, with the long-term goal of creating a complementary venue for probing HEP phenomena \cite{Dalmonte_review,Pasquans_review,Zohar_review,Alexeev_review,aidelsburger2021cold,Zohar_NewReview,klco2021standard,Bauer_review,dimeglio2023quantum,Cheng_review,Halimeh_review}.

\begin{figure}[t]
	\includegraphics[width=0.95\linewidth]{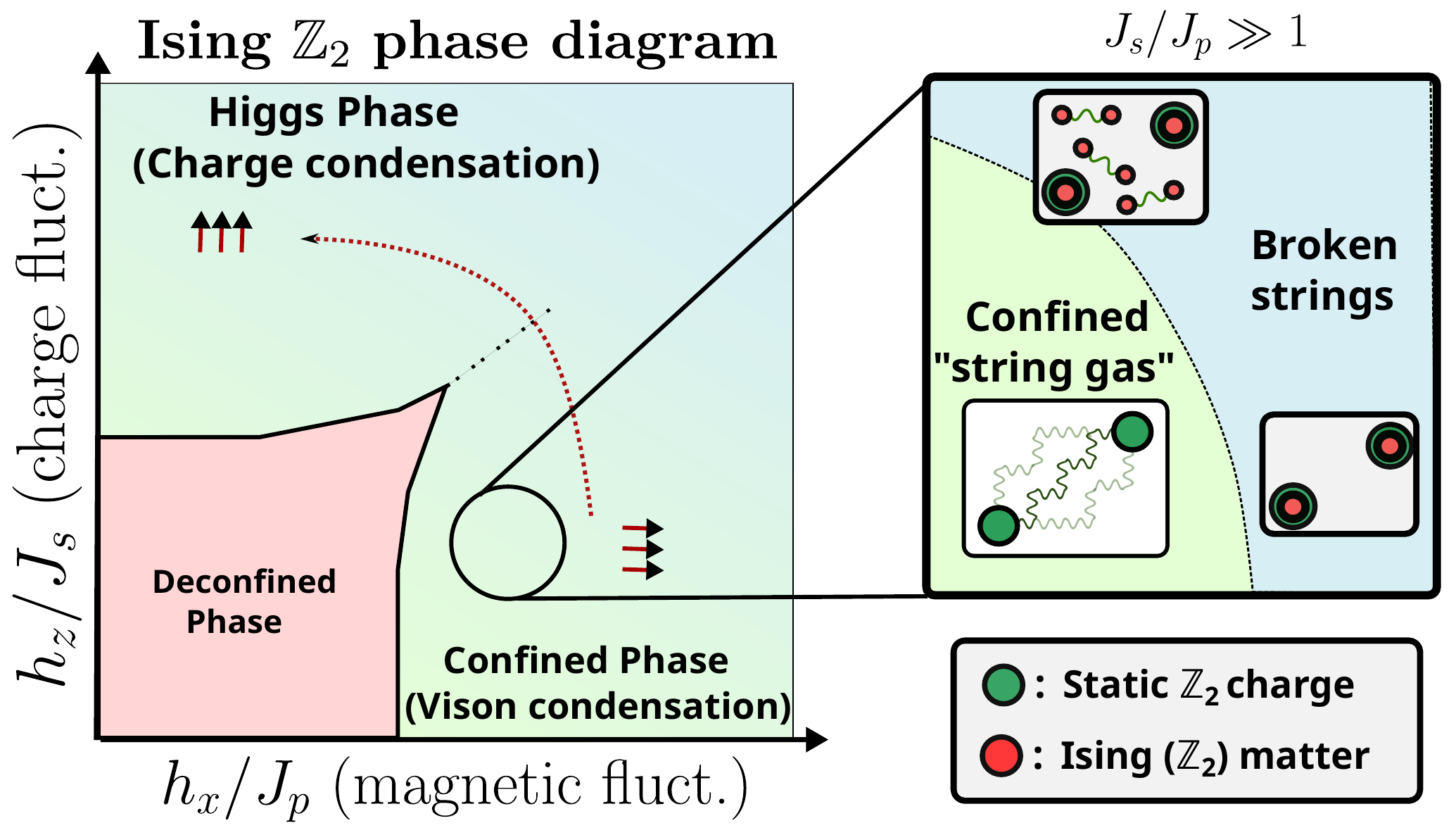}
	\caption{Left: quantum phase diagram of the model described by Eq.~\eqref{eq:2DFS}. The general characterization of the phases does not change when two static charges are present, but we can now distinguish different regimes depending on whether $\Zt$ electric strings are broken. Right: zooming in on the confined region of the phase diagram, we can distinguish different regimes depending on the behavior of strings. As described in the main text, string breaking can be driven by increasing the strength of either external field, but also by decreasing the strength of the magnetic field. 
	\label{fig:phase_diagram}}
\end{figure}

In particular, $\mathbb{Z}_2$ LGT \cite{Wegner1971} has emerged as a paradigmatic model for studying phenomena relevant to HEP, condensed matter, and quantum many-body physics. When coupled with dynamical matter, the Ising gauge theory recently shed new light on various physical phenomena such as confinement \cite{Borla2019,PhysRevX.10.041007,kebric2021confinement,homeier2023,Fromm2024,Kebric2024FiniteTconfinement,Linsel2024}, exotic phase transitions \cite{Grover2016,Gazit2018, konig2019soluble, PhysRevX.11.041008,  Borla_doconfined, xu2024critical}, the Higgs mechanism \cite{Borla2020, verresen2022higgs} and
constrained quantum dynamics \cite{Iadecola2020quantum,aramthottil2022scar,desaules2024massassistedlocaldeconfinementconfined}. Importantly, $\mathbb{Z}_2$ LGTs have recently been experimentally realized on quantum simulators in one \cite{Mildenberger2025,de2024observationstringbreakingdynamicsquantum} and two \cite{cochran2024visualizingdynamicschargesstrings,gyawali2024observationdisorderfreelocalizationefficient} spatial dimensions. Of particular current interest in quantum simulation experiments of $\mathbb{Z}_2$ LGTs is string-breaking dynamics \cite{de2024observationstringbreakingdynamicsquantum,cochran2024visualizingdynamicschargesstrings}, which has also recently been experimentally observed in $\mathrm{U}(1)$ LGTs \cite{gonzalezcuadra2024observationstringbreaking2,liu2024stringbreakingmechanismlattice,crippa2024analysisconfinementstring2}. String breaking is a paradigmatic HEP phenomenon intimately related to confinement in QCD \cite{Gross2021,Berges_review}. Upon spatially separating two quarks, the energy stored in the string connecting them becomes large enough to initiate quark-pair creation between them, thereby breaking the string. It is worth noting that string breaking is also a topic of significant interest in condensed matter \cite{Verdel2020,Verdel2023,mallick2024stringbreakingdynamicsising}.

In the \texttt{Google Quantum AI} experiment \cite{cochran2024visualizingdynamicschargesstrings}, a $2+1$D $\mathbb{Z}_2$ LGT, equivalent to the toric code with two external fields \cite{Fradkin1979,Trebst2007breakdown,Vidal2009low-energy,Wu2012phase}, was used to study string dynamics on a superconducting-qubit quantum computer. 
By fixing the star and plaquette terms to equal strength, they probed the dynamics of strings formed by creating spatially separated local excitations on top of a prepared ground state. Depending on the degree of confinement manifest in the strength of one of the external field terms, three different regimes of string dynamics were observed. Whereas in the deconfined phase the string is quite dynamic and fluctuates all over the lattice, in the confined phase a more nuanced picture emerges. At weak confinement, the string fluctuations in the transverse direction are strong, while at larger confinement, they are frozen.

In this Letter, we employ the density matrix renormalization group technique (DMRG) \cite{Uli_review} to investigate string breaking in this $2+1$D $\mathbb{Z}_2$ LGT deep within the confined phase; see Fig.~\ref{fig:phase_diagram}. In this regime, where charges are connected by strings of minimal length, we provide a detailed account of string-breaking phenomena that result from a number of different mechanisms. We also show how the presence of a weak magnetic field allows an exact mapping to hopping fermions on an open one-dimensional chain.

\textbf{\emph{Model.---}}We focus on the $2+1$D $\mathbb{Z}_2$ LGT on a square lattice, which is the model realized in the recent \texttt{Google Quantum AI} experiment \cite{cochran2024visualizingdynamicschargesstrings}. It is described by the Hamiltonian \cite{Fradkin1979}
 
\begin{align}\nonumber
\hat{H}=&-J_s\sum_\rr \hat\tau^z_\rr-J_p\sum_{\rr^*} \hat B_{\rr^*}\\
&-h_z \sum_{\rr,\eta} \hat\tau^x_\rr \hat\sigma^z_{\rr,\eta} \hat\tau^x_{\rr+\eta}-h_x\sum_{\rr,\eta} \hat\sigma^x_{\rr,\eta},
\label{eq:2DFS}
\end{align}
where $\hat{\tau}^{z}_\mathbf{r}$ represents the matter particle-number operator on site $\mathbf{r}$ with $J_s$ the chemical potential, and $\hat\sigma^{x(z)}_{\rr,\eta}$ is the electric (gauge) field operator on the link emanating from site $\mathbf{r}$ in the direction $\eta$, with $h_x$ the electric-field strength. The plaquette operator $\hat B_{\rr^*}=\prod_{b\in \square_{\rr^*}} \hat\sigma^z_b$ with strength $J_p$, where the index $\rr^*$ labels the sites of a dual lattice formed by the centers of the plaquettes. The three-body term in Hamiltonian~\eqref{eq:2DFS} describes gauge-matter coupling with strength $h_z$. We also define the star (or vertex) operator $\hat A_{\rr}=\prod_{\eta \in +_\rr}\hat\sigma^x_{\rr, \eta}$, the product of $\hat\sigma^x$ on the four links connecting at the vertex $\rr$.
The Hamiltonian \eqref{eq:2DFS} is invariant under the set of local gauge transformations $\hat{G}_\rr = \hat A_{\rr}\hat\tau^z_{\rr}$,
which relates the $\Zt$ electric lines, i.e., links where $\sigma^x=-1$, emanating from a site to the $\Zt$ charge on the same site. Physical states satisfy Gauss's law: $\hat{G}_\rr|\psi\rangle=Q_\rr |\psi \rangle$, where $Q_\rr={\pm} 1$ denotes the absence or presence of a static background $\Zt$ charge on that particular site. The Hilbert space separates into an exponential number of sectors, each determined by a different distribution of background charges. In order to satisfy Gauss's law, a basis of physical gauge-invariant states must be formed by connecting $\Zt$ charges, either dynamical or static, with electric strings. In the following, we will focus on the case where there are two static charges at sites $\rr_1$ and $\rr_2$ ($Q_{\rr_1}=Q_{\rr_2}=-1$, $Q_{\rr\neq\rr_1,\rr_2}=+1$), and study the properties of the ground state and of the strings connecting them in different regimes.

By fixing the gauge sector, matter degrees of freedom can be integrated out of Hamiltonian \eqref{eq:2DFS}. The model then reduces to the toric code in a tilted field \cite{tupitsyn2010}
\begin{align}\nonumber
\hat{H}=&-J_s\sum_\rr Q_\rr \hat A_\rr- J_p\sum_{\rr^*} \hat B_{\rr^*}\\
&-h_z \sum_{\rr,\eta} \hat\sigma^z_{\rr,\eta}-h_x\sum_{\rr,\eta} \hat\sigma^x_{\rr,\eta},
\label{eq:tc_field}
\end{align}
which involves only gauge-invariant link variables. The first term, which includes the static charges, is obtained by solving Gauss's law to yield $\hat\tau^z_\rr = Q_\rr \hat A_\rr$. While it has been appreciated recently that the models \eqref{eq:2DFS} and \eqref{eq:tc_field} differ in some aspects that involve physical boundaries and entanglement cuts \cite{verresen2022higgs, PhysRevX.15.011001}, in this paper we do not investigate such aspects.
For all our numerical simulations, we use Hamiltonian~\eqref{eq:tc_field}. 

In the absence of the external fields $h_{x,z}$, Hamiltonians \eqref{eq:2DFS} and \eqref{eq:tc_field} are invariant under magnetic (electric) one-form symmetries generated by Wilson ('t Hooft) loops. Additionally, the model exhibits an electric-magnetic duality that amounts to exchanging the roles of the star
and plaquette operators and redefining the couplings accordingly. For $J_s=J_p=1$, this results in the phase diagram in Fig.~\ref{fig:phase_diagram} being symmetric under reflections with respect to the main diagonal. We note that the placement of static charges breaks this duality, unless one also flips the sign of $J_p$ on the relevant plaquettes.

The ground-state phase diagram, in the absence of external charges, is well known. A detailed analysis is already present in the seminal paper of Fradkin and Shenker \cite{Fradkin1979}, while subsequent studies using Monte Carlo \cite{tupitsyn2010, PhysRevB.85.195104} and tensor network \cite{xu2024critical} techniques highlighted a number of more subtle properties in the vicinity of the critical points. As shown in Fig.~\ref{fig:phase_diagram}, there are only two distinct bulk phases. For low values of the external fields, the gauge fields are deconfined, both charge and magnetic fluctuations are inhibited, and the ground state approaches the one of the toric code, exhibiting $\Zt$ topological order. At strong external fields, on the other hand, there is a single phase of matter as large values of $h_x$ and $h_z$ force a paramagnetic state along a direction determined by their relative strength. In physical terms, however, it is useful to identify two different regimes which are only separated by a smooth crossover. In the electrically dominated phase charges are confined and magnetic excitations (visons) condense. For fixed ``chemical potential'' $J_s$ and large $h_x$ the ground state has no electric strings and no dynamical $\Zt$ charges. If static charges are present, these are neutralized by dynamical ones to maintain the no-string condition. In the Higgs regime (large $h_z$), on the other hand, visons are confined and charges condense.

\begin{figure}[t]
	\includegraphics[width=0.9\linewidth]{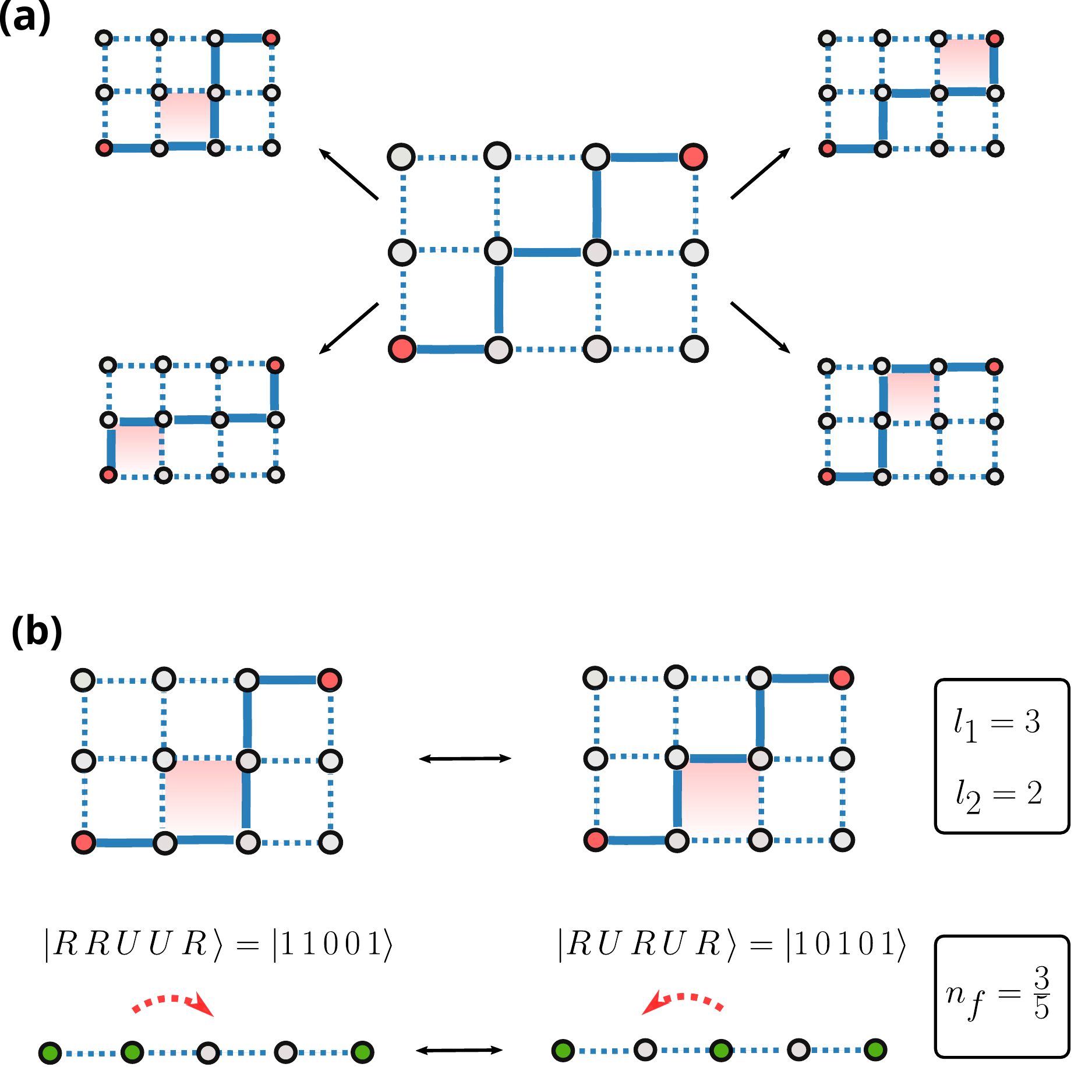}
	\caption{(a) On a $3\times2$ rectangular patch with static charges at the opposite corners (sites marked in red), $10$ string configurations of minimal length $l=5$ are possible. At the center, we show the string with the maximum number of corners. In the presence of a plaquette term, this string can resonate with the four other displayed string configurations. Transitions are caused by the action of the operator $\hat B_{\rr^*}$ on each of the shaded plaquettes.
    (b) Mapping between a system of shortest strings and one-dimensional fermions. The transitions between string configurations that are connected by a single application of $\hat B_{\rr^*}$ correspond to the nearest-neighbor hopping of a fermion. The fermion filling is fixed to $l_1/(l_1+l_2)$.}
	\label{fig:resonating_strings} 
\end{figure}

In the absence of dynamical matter, the model reduces to a pure $\Zt$ LGT. This limit can be obtained by suppressing matter fluctuations ($h_z=0$) and by penalizing occupancy with a large ``chemical potential'' $J_s$.
In this scenario the potential between static $\Zt$ charges is known \cite{Fradkin2013}. In the deconfined phase, the charges are expected to attract each other via an exponentially decaying potential. 
In the confined phase, on the other hand, the attractive potential grows linearly with the distance and is equal to the energy cost of the shortest possible string connecting the charges. At the quantum critical point separating the two phases, which lies in the Ising* universality class, the interaction potential is expected to follow a $1/R$ power-law behavior \cite{peskin1980critical}. Once dynamical matter is introduced the potential between charges can be screened by $\Zt$ charged particles, leading to string breaking phenomena. In the limit of infinite string tension $h_x \rightarrow \infty$, particles will occupy the sites where the static charges are to completely neutralize them.

\textbf{\emph{String phenomenology in the confined phase.---}}We now focus on the phenomenology of strings in the confined phase, $h_x/J_p\gg h_z/J_s$. Sufficiently deep in the confined phase, the electric lines connecting the two static charges have a high energy cost proportional to their length, and must therefore be as short as possible. On a square lattice, unless the two charges are in-line, the shortest possible string connecting them is not unique and, in general, the ground state of the system consists of a superposition of all possible shortest strings.

In the presence of a weak magnetic term, the actual ground-state configuration can be determined by regarding $J_p$ as a small perturbation. Whenever a string forms a corner, it can resonate with another string of the same length through the application of a plaquette operator $\hat{B}_{\rr^*}$ on that specific corner; see Fig.~\ref{fig:resonating_strings}. Since the resonance will lower the ground state energy, strings with the largest number of resonances should appear with a larger weight in the superposition. As the number of available resonances corresponds to the number of corners, the perturbation favors the formation of strings with a ``zigzag'' structure. Given a rectangular patch of size $l_1 \times l_2$, each string configuration connecting the bottom left and top right corners can be obtained by moving horizontally $l_1$ times and vertically $l_2$ times, in any order. The string configurations are therefore in one-to-one correspondence with the ${l_1+l_2 \choose l_1} = {l_1+l_2 \choose l_2}$ permutations of the tuple
\begin{equation}
    (\underbrace{1,1,\dots 1}_{l_1 \text{ times}},\underbrace{0,0, \dots 0}_{l_2\text{ times}}),
\end{equation}
where $1$ represents a horizontal move and $0$ a vertical move. In this framework, transitions between strings correspond to swaps of neighboring pairs of $1$ and $0$, i.e., a southeast corner represented by $(0,1)$ is changed to a northwest corner represented by $(1,0)$. But this is exactly the behavior of $N=l_1$ nearest-neighbor-hopping fermions on an open chain of length $L=l_1+l_2$, with $1$ labeling a site occupied by a fermion and $0$ an empty site. Swapping the roles of $l_1$ and $l_2$ would correspond to simulating the equivalent system of holes, with identical results due to particle-hole symmetry. The plaquette strength $J_p$ plays the role of hopping parameter. 

What is the leading effect of a weak matter coupling $h_z$?
It is straightforward to demonstrate that the matter coupling $h_z$ does not split the energies of the shortest strings at second order in perturbation theory. At fourth order, it generates the plaquette term perturbatively, which leads to renormalization of the fermion hopping.

As a result of this mapping, deep in the confined regime the ground state of the string is a free Fermi gas of one-dimensional fermions occupying an open chain of length $L$ at the filling fraction $l_1/(l_1+l_2)$.
This picture also allows us to determine string excitations. These correspond, in fact, to the familiar particle-hole pairs in the Fermi gas. 

\textbf{\emph{Numerical study of string breaking.---}}In this section we present the results of numerical DMRG simulations carried out with the tensor network \texttt{Python} library \texttt{TeNPy} \cite{Hauschild2018, 10.21468/SciPostPhysCodeb.41}. This method returns a controlled approximation of the ground state of the system in the form of a matrix product state (MPS) \cite{Uli_review}. While the algorithm was originally devised to handle $1+1$D systems, it is readily adapted to $2+1$D cylindrical geometries by defining an ordering of sites and treating the cylinder as a winding chain. This has the drawback of artificially introducing long-range interactions, which cause an exponential growth in the virtual dimension of the MPS needed to reach a given accuracy. This limits the circumference sizes that can be reliably simulated. To make our analysis as close as possible to the $2+1$D limit while retaining good numerical accuracy, we consider a cylinder with $L_x=30$ and $L_y=6$, and focus on its central section where the two charges are placed at the opposite corners of a patch of size $l_1 \times l_2$, with $l_1=5$ and $l_2=3$. This mitigates finite-size effects. For this setup, we scan different ranges of parameters to unveil all possible mechanisms that lead to string breaking, and to clarify the physical role of the three independent couplings $h_x$, $h_z$ and $J_p$.

\begin{figure}[t]
\centering
    \captionsetup[subfigure]{labelformat=empty}
    \subfloat[\label{subfig:hx_breaking}]{}    \subfloat[\label{subfig:hz_breaking}]{}
	\includegraphics[width=0.95\linewidth]{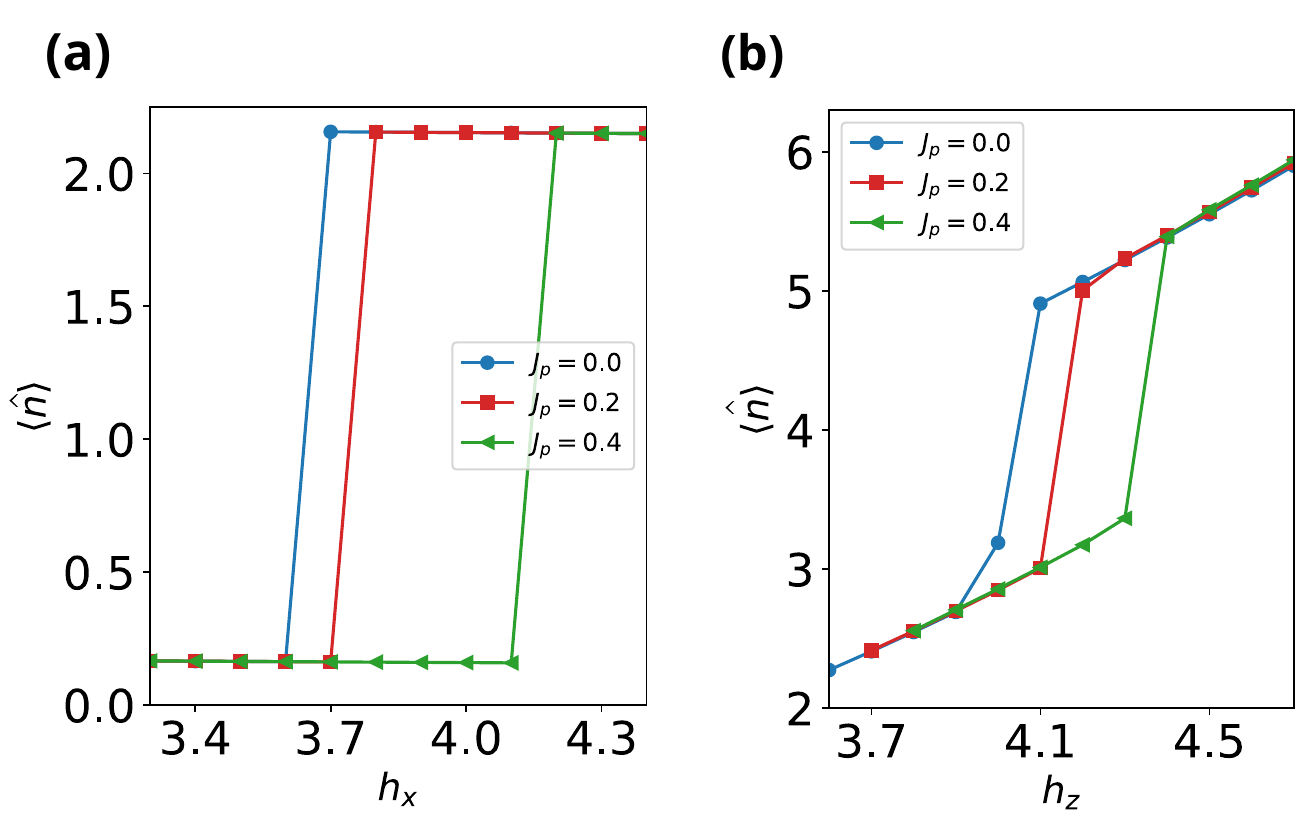}
	\caption{String breaking driven by directly increasing the energy cost of the (a) electric string and (b) charge (pair) fluctuations. As a figure of merit, we compute the total number of particles in the system, which experiences a sharp jump $\Delta n=2$ when a pair forms and neutralizes the static background. In both cases, a finite magnetic coupling $J_p$ stabilizes the strings, shifting the breaking point to the right. The numerical simulations are performed on cylinders of length $L_x=30$ and circumference $L_y=6$. The charges are placed in the central section to avoid boundary effects, at a horizontal distance $l_x=5$ and vertical distance $l_y=3$. The star coupling is $J_s=15$, while we fix $h_z=1$ and $h_x=3$ in panels (a) and (b), respectively.} 
	\label{fig:hx-hz-break} 
\end{figure}

\begin{figure}[t]
\centering
	\includegraphics[width=\linewidth]{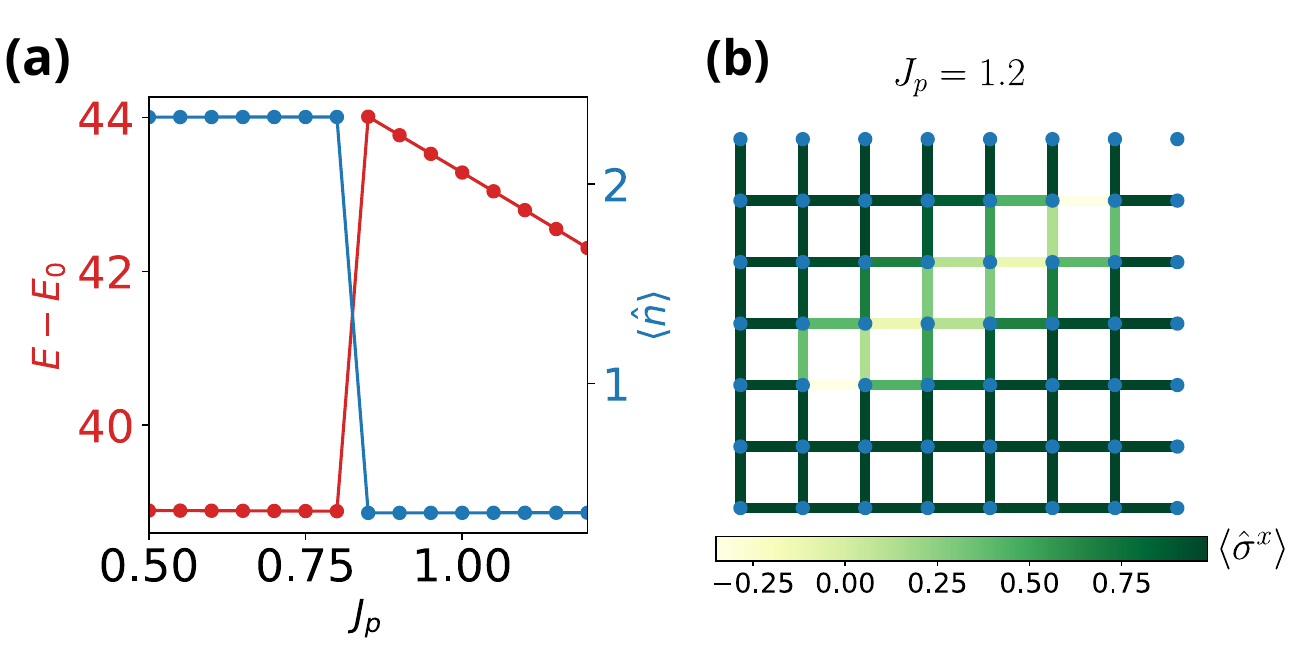}
	\caption{String breaking can be driven by decreasing the plaquette strength $J_p$. (a) The string breaking point is signaled by a sudden jump in the energy and particle number, caused by the creation of two charges that neutralize the static $\Zt$ background, which lowers the ground-state energy. (b) Expectation values of $\hat{\sigma}^x$ in the unbroken string phase. One can see that the two charges are connected by a superposition of strings. The pattern predicted in the main text is clearly recognizable, with the zigzag strings having a higher weight in the superposition. The simulation is done on a cylinder of length $L_x=30$ and circumference $L_y=6$, for $h_x=3$, $h_z=1$, and $J_s=10$.}
	\label{fig:Jp-break} 
\end{figure}

 At $J_p=h_z=0$, the Hamiltonian is classical and whether unbroken strings exist depends solely on the energetic competition between the star and electric terms. The cost of a string connecting the two charges is $\Delta E_{s} = 2 h_x l$, where $l=l_1+l_2$ is the total length of the string. On the other hand, a broken string comes at the cost $\Delta E_{b} = 4 J_s$ of creating two particles that neutralize static charges. In this simple scenario, therefore, string breaking occurs once the critical string tension $h_x^* = 2 J_s/l$ is reached. This is also confirmed in the presence of small quantum fluctuations generated by the plaquette term, as shown in Fig.~\ref{subfig:hx_breaking}.

Next, we consider the possibility of triggering string breaking by increasing the strength of particle fluctuations, determined by the $h_z$ field. To test this, we start from the same configuration as above with stable unbroken strings and increase $h_z$.  As shown in Fig.~\ref{subfig:hz_breaking}, the strings break for large values of $h_z$. In contrast to the electric coupling-driven breaking, this phase is characterized by strong charge fluctuations. Indeed, larger values of $h_z$ facilitate matter creation, thereby destabilizing strings.

Finally, it is important to understand how a finite plaquette term affects the string breaking. Interestingly, as shown in Fig.~\ref{fig:hx-hz-break}, we find that the magnetic term has a stabilizing effect on the strings, leading to an increase in the critical values $h_x^*$ and $h_z^*$ of the external fields at which strings break. This makes sense because, as described above, $J_p$ is responsible for string resonances that lower the energy of the system. Following the same logic, we expect that string breaking can occur by \textit{lowering} the plaquette coupling. This is confirmed by the parameter scan presented in Fig.~\ref{fig:Jp-break}, which shows a broken-string regime for low values of the coupling $J_p$. Starting at large values of $J_p$, we find that the ground-state energy of the system with static charges steadily increases with respect to that in the homogeneous sector where no static charges are present ($Q_\rr=+1,\,\forall\rr$). However, below a critical value $J_p^*$, the string breaks, indicated by the sudden jump in matter occupation by $2$, and the ground-state energy difference is lowered to a constant value.

\textbf{\emph{Summary and outlook.---}}In this Letter, we presented an in-depth study of ground-state string behavior in a $\Zt$ lattice gauge theory with Ising matter in the presence of two static charges. We outlined the role of the plaquette term in stabilizing strings in the confined phase. We provided a mapping to a free-fermionic chain in the presence of a weak magnetic field. Our findings can be readily tested on superconducting-qubit quantum computers such as the one used in the experiment~\cite{cochran2024visualizingdynamicschargesstrings}. Several possible generalizations will be the subject of future research. Novel phenomena can arise when considering different types of matter fields, such as fermions, or models with different global or gauge symmetries. Different lattice geometries may also have a significant impact on the nature of string breaking, especially those relevant to Rydberg-atom realizations \cite{homeier2023}.

{\footnotesize\textbf{\emph{Acknowledgments.---}}U.B.~acknowledges support from the Israel Academy of Sciences and Humanities through the Excellence Fellowship for International Postdoctoral Researchers. S.M.~is supported by Vetenskapsr{\aa}det (grant number 2021-03685) and acknowledges support provided by Nordita. J.C.H.~acknowledges funding by the Max Planck Society, the Deutsche Forschungsgemeinschaft (DFG, German Research Foundation) under Germany’s Excellence Strategy – EXC-2111 – 390814868, and the European Research Council (ERC) under the European Union’s Horizon Europe research and innovation program (Grant Agreement No.~101165667)—ERC Starting Grant QuSiGauge. This work is part of the Quantum Computing for High-Energy Physics (QC4HEP) working group.}


%

\end{document}